# On the reliability of linear band structure methods

G. Kontrym-Sznajd[1], M. Samsel-Czekała[1], G.E. Grechnev[2], and H. Sormann[3]

[1] *Institute of Low Temperature and Structure Research, Polish Academy of Sciences, P.O. Box 1410, 50-950 Wrocław 2, Poland*
[2] *B. Verkin Institute of Low Temperature Physics and Engineering, 47 Lenin Avenue, Khar'kov 310164, Ukraine*
[3] *Institut für Theoretische Physik-Computational Physics, Technische Universität Graz, Petersgasse 16, A- 8010 Graz, Austria*

**Abstract.** - We discuss an efficiency of various band structure algorithms in determining the Fermi surface (FS) of the paramagnetic $ErGa_3$. The linear muffin-tin orbital (LMTO) in the atomic sphere approximation (ASA) method and three full potential (FP) codes: FP-LMTO, FP linear augmented plane wave (FLAPW), and FP local orbitals (FPLO) methods are employed. Results are compared with electron-positron (e-p) momentum densities reconstructed from two dimensional angular correlation of annihilation radiation (2D ACAR). Unexpectedly, none of used modern FP codes is able to give satisfying description of the experimental data that are in perfect agreement with LMTO-ASA results. We suspect that it can be connected with a different choice of the linearization energy.

## 1 Introduction

The rare-earth (RE) compounds constitute an interesting subject of investigations due to their magnetic properties [1]. Lanthanides model assumes that the magnetic coupling between the 4*f* electrons is carried out indirectly via polarization of conducting electrons, according to the Ruderman-Kittel-Kasuya-Yosida (RKKY) theory [2] and that magnetic properties are controlled by the FS topology (so-called nesting – flat FS sheets) [3]. The first direct observation of such an FS feature was possible thanks to the study of 2D ACAR spectra [4-10]. In this unique technique one measures line projections of the e-p pairs in the extended momentum space $\rho(\mathbf{p})$. From such spectra for different directions of projections one can reconstruct $\rho(\mathbf{p})$ in 3D momentum space applying tomography techniques [11]. Next, using the LCW folding [12], i.e., a conversion from the extended $\mathbf{p}$ into reduced $\mathbf{k}$ space, one may reproduce the FS.

In the case of ErGa$_3$ 2D-ACAR experiment was carried out in the paramagnetic state at temperature of 60 K and in the magnetic field 1.8 T (for more details see [9]). Four projections were measured and next deconvoluted employing the Van Citter-Gerhardt algorithm [13]. The e-p momentum densities $\rho(p)$ were reconstructed by Cormack's method (CM) [14] and the lattice harmonics expansion (LHE) algorithm [15]. Finally, the 3D-LCW transformation [12] was applied to get $\rho(k)$ (occupation numbers "seen" by positrons). These results were compared with band structure calculations performed via four different methods: LMTO-ASA [16] and three modern codes containing FP instead of ASA: FP-LMTO [17], FLAPW [18], and FPLO [19].

## 2  RESULTS

Primary band-structure calculations for ErGa$_3$, crystallizing in the cubic AuCu$_3$–type structure with the lattice constants a=4.212 Å, were performed using the LMTO-ASA code [16]. Thus obtained FS was in very good agreement with the density $\rho(k)$ reconstructed, via the CM, from 2D ACAR spectra [9], displayed in fig. 1(c) and (a) (the same as in fig. 2 in ref. [9]). The electron-like FS in the 7$^{th}$ band, determined from the reconstructed density, exhibits a nesting feature with a spanning vector in the [110]-type directions, consistent with the overall magnetic structure of the system [20-21] that orders antiferromagnetically at $T_N$ =2.8 K within an incommensurate magnetic structure.

In order to analyze experimental e-p densities in the extended $p$ space new theoretical calculations were performed by using the FLAPW method [18]. They were done both in the extended and reduced zone scheme. Unexpectedly, results in the plane XΓXM, shown in Figs. 1(d) are quite different from both previous LMTO-ASA calculations and reconstructed densities. Due to this reason a reconstruction procedure was carried out once again employing our newest LHE algorithm (it confirmed the former CM results in detail – see fig. 1) as well as we repeated band-structure calculations using another FP codes: FP-LMTO and FPLO.

In the framework of all applied methods we assumed that all eleven 4$f$ electrons in ErGa$_3$ were core states. This can be justified by inelastic neutron scattering experiments [21], which revealed the Γ$_7$ doublet being in ErGa$_3$ the ground state of the 4f multiplet, due to the crystal field splitting. Also, according to the photoemission data [22] the 4f spectral density for Er and Er-based compounds was about 5 eV below the Fermi energy (E$_F$). Moreover, reconstructed densities, drawn in the extended zone, clearly show that there are twelve conduction electrons in ErGa$_3$. Unfortunately, treating 4$f$ electrons as core states we were not able to receive self-consistent results within the FPLO method in any standard way. It was possible both for the previous LMTO-ASA, recent FP-LMTO as well as FLAPW calculations where the 4$f$ states were considered as spin-polarized outer-core wave-functions contributing to the total spin density. The spin occupation numbers of the 4$f$ electrons were fixed by applying the Russel-Saunders coupling scheme to the 4$f$ shell, according to ref. [23]. In this way the shell is not allowed to hybridize with conduction electrons but provides some kind of exchange field that effects the conduction band. This approach corresponds to the standard rare-earth model [24] within the limits of infinite Hubbard repulsion U in the *ab initio* local spin density LSDA [25] scheme for the exchange-correlation effects. So, we obtained localized spin-polarized 4$f$ states, and this treatment has nothing to do with the so-called "LSDA+U" approach.

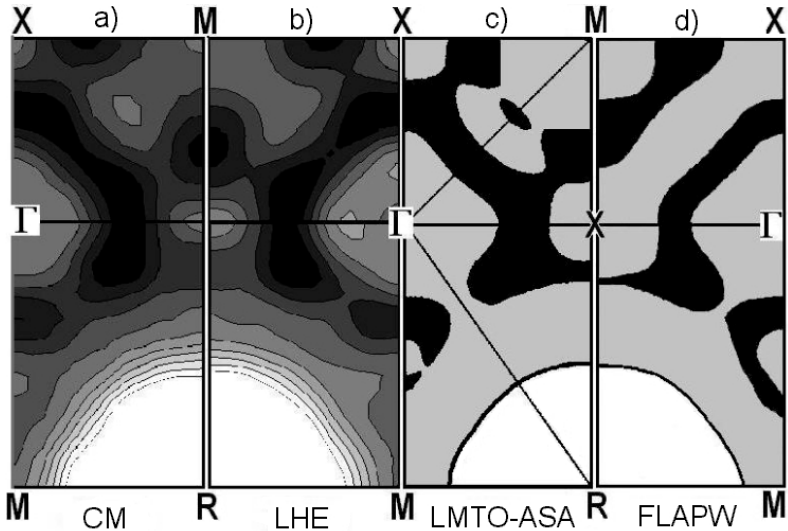

Fig. 1 - Densities $\rho(\mathbf{k})$ in ErGa$_3$ on high symmetry planes, reconstructed by the CM [9] and LHE method, compared with the FS sections calculated via LMTO-ASA [16] and FLAPW codes with spin-polarized 4$f$ core states. The white region centered at the R point contains the occupied states from the 7$^{th}$ valence band and the black area denotes unoccupied states from the 6$^{th}$ band.

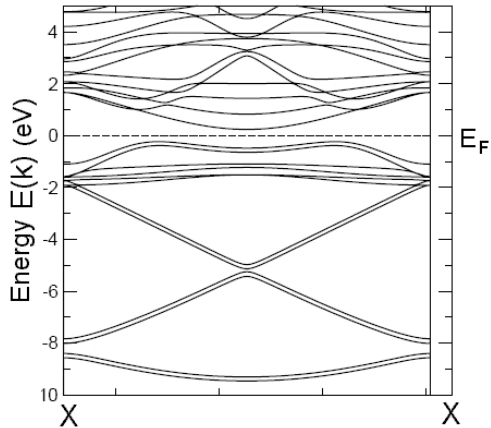

Fig. 2 - The band energies E($\mathbf{k}$) of paramagnetic ErGa$_3$, along the X-X line in BZ, calculated using the FP-LMTO method with spin-polarized 4$f$ core states.

The calculations were performed for the induced paramagnetic configurations of valence electrons within both local spin density (LSDA) – results are displayed in fig. 2 - or generalized gradient (GGA) approximations of the exchange-correlation potential. For the core charge density the Dirac equation was solved self-consistently and then the spin-orbit coupling term was included in the Hamiltonian at each variational step [24,26].

However, as shown in both fig.1 and fig.2, none of these FP algorithms is able to reproduce four Fermi cuts along the X-X line observed in the experimental 2D ACAR data being in excellent agreement with the LMTO-ASA band structure calculations [9].

Why does the LMTO-ASA method produce better results than FP codes? It is seen in fig. 2 that along the X-X line band energies obtained by the FP-LMTO (the same for the FLAPW results not displayed here) are lower at about 0.1 eV from the Fermi energy $E_F$, which is basically of the order of an LSDA-calculation error. Therefore, a lack of the FS cuts along this line obtained by FP algorithms may occur due to this error. However, because it was obtained for both FLAPW and FP-LMTO, it seems that the most possible reason for the difference between the LMTO-ASA and FP results is due to different fixing of the linearization energy $E_{vl}$, around which the expansion of energy dependent wave function is performed. $E_{vl}$ is usually taken to be the center of gravity of the occupied part of the $l$ band. In general, this choice provides more accurate charge densities, magnetic and ground state properties. However, as it was shown by Skriver with his LMTO-ASA implementation [27] another possible choice, $E_{vl} = E_F$, presumably gives more accurate FS. This approach was applied to previous LMTO-ASA calculations for $ErGa_3$ [16] but not to our present FP calculations, since the standard FP codes have not such an option.

## 3 CONCLUSIONS

The Fermi surface of paramagnetic $ErGa_3$ was probed by various band-structure codes within the density functional theory (DFT) and compared with densities $\rho(k)$ reconstructed from 2D ACAR spectra. Surprisingly, none of the used advanced FP codes is able to reproduce the experimental results which agree very well with the former LMTO-ASA results. It can be an evidence of some failure in construction of an atomic potential in RE or an inefficient choice of internal parameters, presumably the linearization energy. More additional calculations within FP methods, also with fixing linearization energies at $E_F$, are required to shed light on this problem.

**Acknowledgements** We are very grateful to dr M. Biasini for making available his experimental 2D ACAR data for ErGa$_3$ and to the Polish State Committee for Scientific Research (Grant No. 2 P03B 012 25) for the financial support.